# Coherent cavity-enhanced dual-comb spectroscopy

Adam J. Fleisher,[1,*] David A. Long,[1,3] Zachary D. Reed,[1] Joseph T. Hodges,[1] and David F. Plusquellic[2]

[1]*Material Measurement Laboratory, National Institute of Standards and Technology, 100 Bureau Drive, Gaithersburg, Maryland, 20899, USA*
[2]*Physical Measurement Laboratory, National Institute of Standards and Technology, 325 Broadway, Boulder, Colorado, 80305, USA*
[3]*david.long@nist.gov*
[*]*adam.fleisher@nist.gov*

**Abstract:** Dual-comb spectroscopy allows for the rapid, multiplexed acquisition of high-resolution spectra without the need for moving parts or low-resolution dispersive optics. This method of broadband spectroscopy is most often accomplished via tight phase locking of two mode-locked lasers or via sophisticated signal processing algorithms, and therefore, long integration times of phase coherent signals are difficult to achieve. Here we demonstrate an alternative approach to dual-comb spectroscopy using two phase modulator combs originating from a single continuous-wave laser capable of > 2 hours of coherent real-time averaging. The dual combs were generated by driving the phase modulators with step-recovery diodes where each comb consisted of > 250 teeth with 203 MHz spacing and spanned > 50 GHz region in the near-infrared. The step-recovery diodes are passive devices that provide low-phase-noise harmonics for efficient coupling into an enhancement cavity at picowatt optical powers. With this approach, we demonstrate the sensitivity to simultaneously monitor ambient levels of $CO_2$, CO, HDO, and $H_2O$ in a single spectral region at a maximum acquisition rate of 150 kHz. Robust, compact, low-cost and widely tunable dual-comb systems could enable a network of distributed multiplexed optical sensors.

## 1. Introduction

Dual-comb spectroscopy has shown great potential as a fast, accurate, and high-resolution alternative to existing interferometric methods that require lengthy integration times and precision moving parts (e.g., Fourier transform spectroscopy) [1,2]. The inclusion of long interaction paths between a probe comb and a gas-phase sample, either via an open-air path [3] or an enhancement cavity [4], has yielded impressive detection limits for several small molecules important for precision atmospheric monitoring. In order to achieve appropriate sensitivities, however, those experiments required mode-locked lasers with either high-bandwidth phase-locked loops [5] or sophisticated real time signal processing [6,7] to maintain frequency comb coherence for an integration time of $\geq 1$ s. Without further improvements in long-term coherence and a reduction in instrument complexity to maintain this coherence, dual-comb spectroscopy is unlikely to find widespread acceptance in many areas of applied spectroscopy.

Direct frequency comb spectroscopy using dispersive or Fourier transform methods for broadband detection has found a wide variety of applications in atomic and molecular physics, as well as in industrial, chemical, environmental, and medical sensing [8–10]. For many applications, however, the 100s of nm spectral coverage of mode-locked laser (MLL) based combs is not required and a more robust, frequency-agile platform tailored for desired properties such as scan speed, spectral resolution and tooth power would be most beneficial. Towards this goal, the engineering of optical frequency combs has followed an alternate route to include several novel photonic devices such as various microresonator geometries [11] or cascaded intensity and phase modulators [12], the latter often combined with highly non-linear fibers [13,14]. While comb generation based on electro-optic phase modulators (EOMs) was first demonstrated in the 1990s for metrology [15–17] and interferometry [18], only recently have the advantages of dual-EOM comb systems been realized for spectroscopy [19–22]. In cases where the targeted spectral coverage is narrow, the EOM-based approach can directly overcome several of the technical challenges associated with MLL dual-comb spectroscopy [23]. Specifically, EOM combs offer 1) inherently high relative phase coherence between the two arms of the interferometer owing to its self-heterodyne nature [24,25], and 2) the frequency-agility and precision of the driving low-phase-noise microwave electronics for exquisite control of spectral coverage, resolution, scan speed and down-conversion bandwidth. These features make for an efficient and facile way to couple all comb teeth into narrow resonance modes of an enhancement cavity for high resolution and ultra-sensitive spectroscopic detection.

Multiheterodyne spectroscopy performed using two stabilized frequency combs can retrieve broadband molecular absorption and phase information from an interferogram recorded as a function of the pulse-to-pulse delay between probe and local oscillator (LO) combs. In the time domain, two trains of laser pulses with slightly different pulse delays (equal to the inverse of the frequency comb mode spacing, $t_d = 1/f_{mod}$) are combined on a single photodetector, resulting in interferograms that repeat in time every $T_i = 1/\Delta f_{mod}$, where $\Delta f_{mod} = f_{mod,1} - f_{mod,2}$. In the frequency domain, the individual comb teeth of the probe comb are down-converted into the radiofrequency (RF) domain by mixing with corresponding teeth of the LO. This results in a compression of the optical probe comb bandwidth for detection by the factor $f_{mod}/\Delta f_{mod}$, significantly simplifying the acquisition and analysis of the optical signal.

Importantly, when working with MLLs, the $f_{mod}$ of each comb (analogous to the comb repetition rate $f_{rep}$) is defined by a physical length (i.e., the laser cavity) and is therefore nominally fixed after laser construction. The MLLs must also exhibit exceptionally high stability in order to maintain a relative linewidth less than $\Delta f_{mod}$ (analogous to the difference in the comb repetition rates $\Delta f_{rep}$). Multiple photodetectors or narrow-bandwidth optical filters are therefore required when all of MLL teeth cannot be sufficiently compressed [5,26].

In order to address current challenges in dual-comb spectroscopy pertaining to relative comb coherence and cost-effective, robust instrument development, we demonstrate here dual-EOM comb spectroscopy with a combination of three novel enhancements: 1) the inclusion of step-recovery diodes as passive, in-line, hands-free devices for generating coherent RF pulses, 2) the design of a low-bandwidth phase-locked loop which allowed for > 2 hours of mutual comb coherence without software corrections, and 3) a real time data streaming and time-domain coherent averaging procedure which improved data throughput to nearly 100 %.

## 2. Methodology of comb generation

The near-infrared EOM combs reported here consisted entirely of polarization-maintaining optical fiber components and are illustrated in Fig. 1(a). A continuous-wave (CW) external cavity diode laser (ECDL), tunable from 1570 nm to 1630 nm [27,28], was coupled to an erbium-doped fiber amplifier (EDFA, 180 mW optical power) prior to seeding the EOMs for comb generation and cavity-comb coupling. The probe and LO combs were created using phase modulators of nominally 20 GHz 3-dB RF bandwidth and were driven by separate step-recovery diodes used to imprint RF harmonics upon the CW optical carrier in each arm. The time-domain spectrum of the step-recovery diode output, driven by an RF modulation frequency of $f_{mod} \approx 203$ MHz, is shown in Fig. 1(b). The corresponding power spectrum revealed harmonic generation by the step recovery diode throughout the 14-GHz-wide bandwidth of our spectrum analyzer. Optical compression of the probe comb was performed via the LO comb, generated using an identical step-recovery diode driven by a slightly different modulation frequency of $f_{mod} + \Delta f_{mod}$ (where $\Delta f_{mod} = 200$ kHz) and whose carrier frequency was shifted by an acousto-optic modulator (AOM) driven at $f_{AOM} = 57$ MHz. A train of the resulting time-domain interferograms is also shown in Fig. 1(c), along with its corresponding power spectrum.

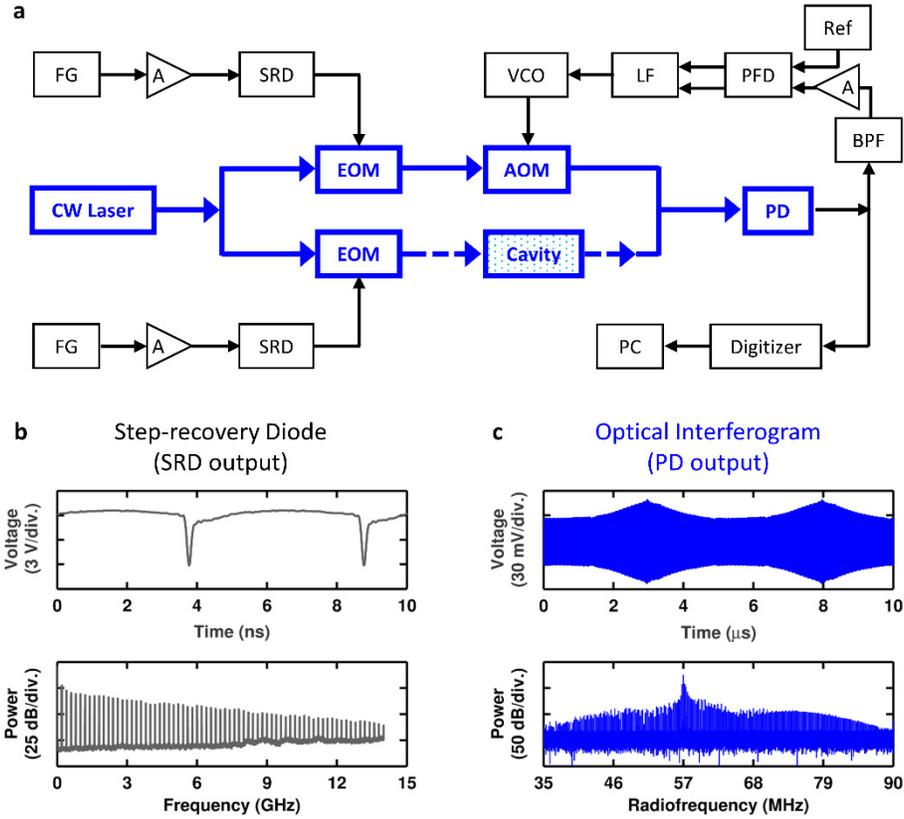

Fig. 1. Block diagram of the dual-comb spectrometer. a, Optical components are illustrated in blue, whereas electronic components are illustrated in black. Fiber optics and free-space laser propagation are illustrated with solid and dashed blue arrows, respectively. Abbreviations used in a are CW, continuous-wave, AOM, acousto-optic modulator, EOM, electro-optic modulator, PD, photodetector, BPF, bandpass filter, A, amplifier, PFD, phase-frequency detector, Ref, reference, LF, loop filter, VCO, voltage-controlled oscillator, FG, function generator, SRD, step-recovery diode, and PC, personal computer. b, c, Time (top) and frequency (bottom) domain spectra of the step-recovery diode (left, gray) and the optical interferogram (right, blue), respectively. In b, the attenuated time trace was sampled using a $4 \times 10^{10}$ s$^{-1}$ oscilloscope with a 4 GHz bandwidth, whereas the attenuated power spectrum was recorded using a frequency-swept spectrum analyzer with a bandwidth of 14 GHz. In c, the time-averaged trace of two interferograms sampled using a $7.5 \times 10^{8}$ s$^{-1}$ digitizer is shown, along with the corresponding Fourier transform of an averaged 4 ms time trace.

As seen in the power spectrum of the train of dual-comb interferograms, the step-recovery diodes generated harmonics up to 26 GHz. This resulted in frequency combs which spanned > 50 GHz of optical bandwidth; an impressive figure-of-merit for such a simple system that utilized only low-cost, passive devices and a single phase modulator per comb. The spectrometer does not require an arbitrary waveform generator, cascaded intensity and phase modulators, or highly non-linear fibers (and corresponding optical amplification) to achieve sufficient spectral broadening. The commercial fiber-coupled lithium niobate EOMs, available to operate at wavelengths from 400 nm to > 2000 nm, had a low $V_\pi = 3.8$ V at a frequency of 1 GHz, which is a significant advancement in modulator technology compared to previous devices that required higher drive powers and optical cavities to produce high-order harmonics [15–18]. Here, the step-recovery diodes were driven at $f_{mod}$ (or $f_{mod} + \Delta f_{mod}$) with an RF power of only 250 mW.

Cavity-comb coupling was maintained via a phase lock of the CW laser carrier frequency, $f_0$, to a given cavity mode using the Pound-Drever-Hall technique [29]. The dual-polarization locking scheme was nearly identical to that described by Long *et al.*, with the exception that demodulation was performed here at $f_{lock} \approx 13$ MHz instead of at twice the PDH modulation frequency (i.e., $2f_{lock}$) [30]. Locking the comb carrier to the stabilized enhancement cavity reduced the relative ECDL laser linewidth by approximately three orders of magnitude to a relative value of 130 Hz [30]. All EOM comb teeth comprising the probe comb were then tuned to be resonant with other given enhancement cavity modes by simply adjusting $f_{mod}$ via a function generator stabilized to a 10 MHz Rb reference clock. The nominal value of $f_{mod}$ was determined by a frequency-agile, rapid scanning (FARS) measurement of the cavity free spectral range ($\approx 203$ MHz) using a high-bandwidth EOM [31]. With $f_{mod}$ referenced to the Rb clock, absolute frequencies on all comb teeth can be determined with $\approx 1 \times 10^{-10}$ relative standard uncertainty (1 s) by measuring the cavity-locked CW laser frequency, $f_0$, relative to an absolute optical reference (i.e., a stabilized, self-referenced MLL frequency comb) [32].

The nominal cavity finesse was measured by cavity ring-down spectroscopy (CRDS) to be $F = \pi/(1 - R_m) = 19\,500$. With a nominal stabilized length of $L = 73.8$ cm and a free spectral range of 203 MHz, the cavity mode linewidth was therefore $\delta_{cav} = 10$ kHz. The mirror transmission and loss (absorption plus scattering) intensity coefficients were estimated from a CW measurement of optical power incident on and transmitted through the enhancement cavity. In the framework of FARS CRDS, this required careful measurement of 1) the optical power in the first-order EOM sideband incident upon the enhancement cavity, 2) the optical power transmitted through the cavity while on resonance, and 3) the cavity mode-matching parameter $\varepsilon$. The mode matching parameter was easily determined to be $\varepsilon = 55.7\,\%$ via a FARS cavity transmission measurement over one free spectral range. Coupled with knowledge of $1 - R_m$ from CRDS, the mirror transmission and mirror losses were successfully deconvolved, where $T_m = 9.2 \times 10^{-5}$ and $\ell_m = 6.9 \times 10^{-5}$, respectively [33].

As with other tightly locked, cavity-enhanced direct frequency comb spectroscopy demonstrations [34], we anticipate that the maximum achievable transmission bandwidth will ultimately be limited by mirror dispersion to $\leq 5$ THz. We note that the enhancement cavity length was actively stabilized to an $I_2$-stabilized HeNe laser with 10 kHz long-term frequency stability, and $f_{mod}$ did not actively track drifts in the cavity free spectral range. At a fictional comb bandwidth of 5 THz and a comb tooth spacing of 203 MHz, the cumulative detuning due to uncompensated cavity length jitter would be $\approx 100$ Hz, much less than $\delta_{cav} = 10$ kHz. Therefore, we anticipate that mirror dispersion would ultimately determine the achievable cavity-coupled bandwidth for broader EOM-based frequency combs.

The probe and LO combs were combined on a fiber-coupled 125 MHz bandwidth photodetector (5 µW average power), and the electronic output of this detector was split into locking and detection signals (see Fig. 1(a)). While the detection signal was sent to a high-speed digitizer, the locking signal was bandpass filtered ($\approx 100$ kHz filter width) around $f_{AOM}$ allowing for the phase of the heterodyne beat signal between the optical carriers of the two combs to be compared to that of a reference signal of frequency equal to a user-defined value of $f_{AOM}$. The outputs of the phase-frequency detector were sent into a loop filter with a proportional-integral corner $\leq 3$ kHz. The loop filter output was used to drive a voltage-controlled oscillator, which provided feedback to the AOM driving frequency $f_{AOM}$.

## 3. Long-term coherence without complexity

By generating two EOM combs originating from the same CW laser, the probe and LO combs that produced the multiheterodyne spectrum exhibited inherently high relative coherence as compared to two free-running MLL combs. This relative coherence, however, was previously limited to $< 1$ s [19] by temperature fluctuations, vibrations, and slow motions of the optical fibers which introduced differential phase noise between the two frequency combs. As discussed above, to reduce this relative phase noise on longer timescales, a low-bandwidth (≤ 3 kHz) RF phase-locked loop (PLL) was implemented which enabled coherent time-domain averaging of interferograms for > 7 200 s. As shown in Fig. 2(a), a 10 s measurement of the frequency comb heterodyne beat signal at 60 MHz revealed that the PLL served to reduce the relative linewidth of the probe and LO combs by more than three orders of magnitude, from ≈ 200 Hz to < 200 mHz.

The multiheterodyne detection signal was digitized with a vertical resolution of 12 bits and at a sampling rate of $7.5 \times 10^8$ s$^{-1}$. At this rate, the data streaming hardware and custom software enabled 99.9 % throughput directly to computer memory via circular stream buffers. A single stream buffer contained $N = 3 \times 10^6$ points (4 ms) and therefore captured a train of 800 interferograms repeating on $T = 5$ μs intervals. Each stream buffer was coadded in real-time using a 14-core PC. A procedure was available for real time jitter/drift correction prior to coadding although not used here since the PLL maintained excellent long-term phase coherence. Successive acquisitions were triggered at $\Delta f_{mod}$ and clocked by a 10 MHz Rb master clock signal. The master clock signal was also used to synchronize all RF sources required to generate each frequency comb. Because the probe comb consisted of only a few hundred comb teeth, a digital lock-in algorithm [35] instead of the usual fast Fourier transform [19] was used to retrieve both the amplitude and phase of all down-converted comb teeth from the time-domain interferograms [20, 21]. Challenges in discrete time-domain sampling like the scalloping of down-converted comb teeth [36] are avoided entirely owing to the frequency-agile, digital control of $f_{mod}$, $\Delta f_{mod}$, and $f_{AOM}$.

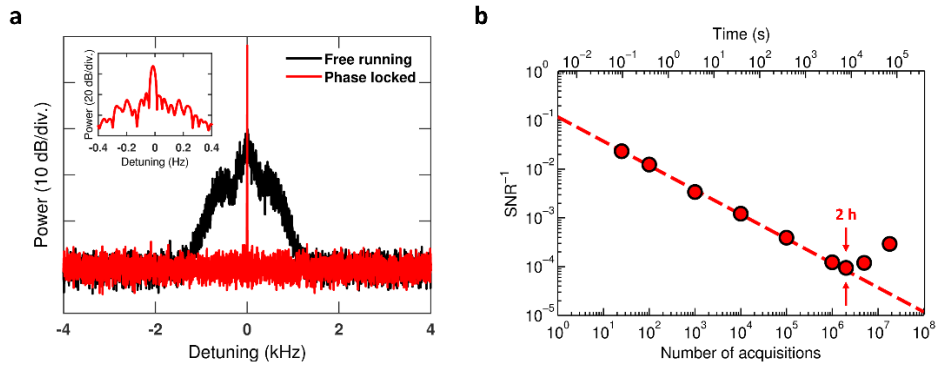

Fig. 2. Coherent averaging of optical interferograms. a, Optical heterodyne beat signal between the 15th harmonics of the probe and local oscillator (LO) combs at a down-converted frequency of 60 MHz. The black trace shows the free-running beating between the two comb teeth, whereas the red trace shows the same beat, but with the AOM phase lock engaged. Both traces were recorded on a spectrum analyzer with 10 s of averaging and a 10 Hz resolution bandwidth (RBW). In the inset, the trace with the AOM phase lock engaged is shown in more detail. This trace was time averaged for 10 s with a 200 mHz RBW. All 257 down-converted comb teeth had a RBW-limited linewidth of 200 mHz when the AOM was phase locked. b, The inverse of the time-domain dual-comb signal-to-noise ratio (SNR) at 60 MHz is plotted versus the number of averaged acquisitions $n$. The acquisition and real-time averaging of successive time traces containing 800 interferograms (4 ms long) was triggered at $\Delta f_{mod}$. The red dashed line represents an increase in SNR proportional to $\sqrt{n}$, the expected trend for coherent signal averaging. The measured SNR$^{-1}$ follows the expected trend for $2 \times 10^6$ averaged acquisitions, or a total time of 2.2 hours as highlighted by the red arrows.

The inverse of the time-domain signal-to-noise ratio (SNR) at a representative down-converted frequency comb tooth (15th harmonic at 60 MHz) versus elapsed time is plotted in Fig. 2(b). For this measurement, the probe comb was redirected around the optical cavity using a fiber-optic switch. Trains of 800 interferograms were acquired and averaged in real time, and the comb tooth amplitude and spectral noise of the averaged time trace were retrieved using a digital lock-in algorithm [35]. The data revealed that time-domain signals could be averaged coherently for > 2 hours without the need for software-based phase corrections. Given that $T = 5$ μs, 2 hours of coherent averaging at a 99.9 % throughput rate resulted in the acquisition of $1.6 \times 10^9$ individual interferograms, or $6 \times 10^{12}$ data points. Without coherent averaging, the transfer, storage, and processing of data at ≈ 6 TB/h would quickly prove to be overwhelming for common computer hardware. Furthermore, the probe comb averaged coherently regardless of whether or not its carrier was locked to the enhancement cavity. This demonstrated that long-term relative coherence does not depend upon the optical comb linewidth [19, 22]. For the latter method, high-bandwidth servos and PLLs [5] or sophisticated signal processing [6, 7] are required to enable coherent averaging for > 1 s.

Long-term (> 1 s) coherent averaging revealed scores of additional frequency comb teeth that would otherwise have been buried in the noise. Plotted in Fig. 3 are the power spectra of two time traces averaged for either $1 \times 10^2$ acquisitions (blue) or $1 \times 10^5$ acquisitions (superimposed in cyan), respectively. If the trend in RF power versus frequency observed for the step-recovery diode (Fig. 1(b), bottom panel) was mapped directly to the magnitude of the EOM optical comb, we estimate that the 130th comb teeth ( ± 26.4 GHz of optical detuning) had exceedingly low optical powers (≈ 100 pW). Given the roll-off in EOM efficiency at higher frequencies (≥ 20 GHz), this estimate represents an upper-bound.

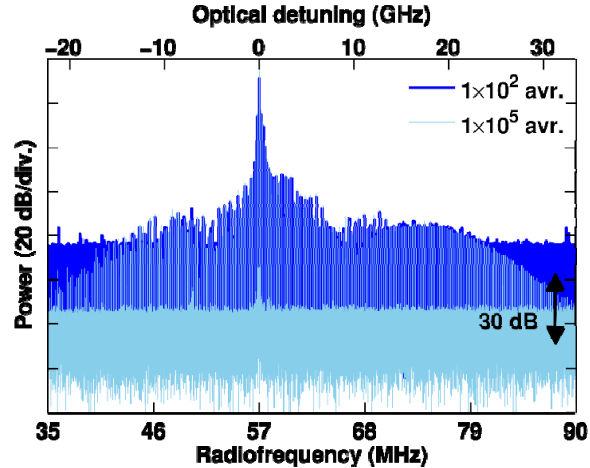

Fig. 3. Greater than 250 comb teeth are revealed. The averaged frequency-domain power spectrum of $1 \times 10^2$ (blue) and $1 \times 10^5$ (cyan) acquired time traces, respectively. Each individual time trace was 4 ms long, and real-time coherent averaging was performed with 99.9 % throughput (duty cycle). A 30 dB reduction in the power spectrum (black double-headed arrow) is expected for an increase in the amplitude SNR proportional to $\sqrt{n}$ .

## 4. Spectral acquisition and normalization

In order to perform cavity-enhanced dual-comb spectroscopy, the cavity transmission must be normalized. Using a fast (< 1 ms) fiber-optic switch, the probe comb could be re-directed through optical fiber around the enhancement cavity and used to normalize the amplitude at each individual comb tooth. The fiber-optic switch was upstream from the enhancement cavity, and could direct the probe comb either through a fiber-to-free-space coupler and cavity-coupling free-space optics or through a secondary fiber optic which led directly to the photodetector and to beating with the LO. The molecular spectrum of synthetic air was acquired by coherent averaging of both probe and reference dual-comb signals. To facilitate the simple integration of the fiber switch software with preexisting acquisition and laser control software, a second digitizer and PC were used to perform molecular spectroscopy.

Each signal was acquired using a 14-bit vertical resolution digitizer at a sampling rate of $8 \times 10^8$ s$^{-1}$ for an acquisition length of 200 μs. A total of 800 acquisitions were recorded in near real-time (98 % duty cycle) and then transferred via USB to the PC for coherent time-domain averaging. After the transfer, the fiber optic switch was triggered, and another 800 acquisitions were recorded to normalize the probe transmission spectrum. The acquisition of both probe and reference combs (320 ms of data) constituted one molecular spectrum. In Fig. 4, the average of 1 000 spectra of a synthetic air sample is shown (see Section 5). Due to lengthy file transfer times from this second digitizer to PC when using the spectral acquisition configuration, one molecular spectrum required ≈ 32 s of total time. With improved custom acquisition software, it will be possible to significantly reduce the dead time between probe and reference comb acquisition to the < 1 ms limit currently set by the fiber optic switch. The comb parameters for cavity-enhanced spectroscopy were $f_{mod}$ = 203.078 689 7 MHz, $\Delta f_{mod}$ = 300 kHz, $f_{AOM}$ = 51 MHz, $M$ = 257 (where $M$ is the number of resolvable comb teeth).

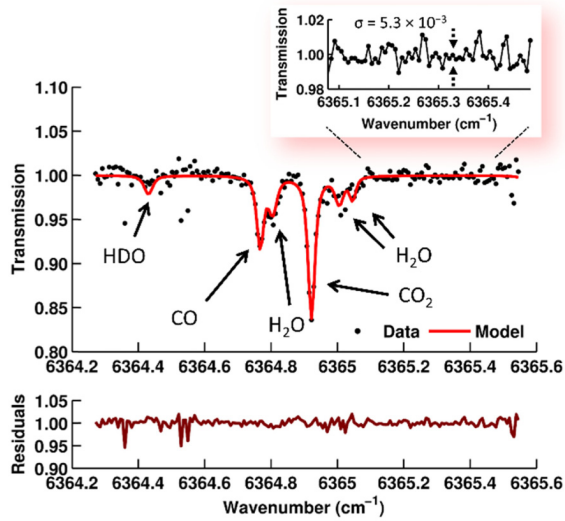

Fig. 4. Dual-comb spectrum of synthetic air. Shown in the top panel is the normalized transmission spectrum of $CO_2$, CO, HDO, and $H_2O$ in air at a total pressure of 13 kPa (black dots, 1 000 averages). A nonlinear least-squares fit (red line) yielded the following mole fractions: $CO_2$, 320(13) μmol/mol, CO, 131(11) μmol/mol, HDO, 12.8(1.5) μmol/mol, and $H_2O$, 4.1(5) %, respectively. The HDO mole fraction and uncertainty is derived from its known isotopic abundance relative to that of $H_2O$ [37]. As seen in the inset, the standard deviation of the spectral noise in areas without molecular absorption was 0.53 %, which corresponds to an intensity absorption coefficient of $\alpha$ = 1.2 × 10$^{-8}$ cm$^{-1}$ in a total integration time of $T$ = 320 s. The average optical power per comb tooth at the photodetector over the entire 12 GHz-wide inset (60 teeth) was ≈ 650 pW. The fit residuals (maroon line) are shown in their entirety in the bottom panel.

## 5. Dual-comb spectroscopy of $CO_2$, CO, HDO, and $H_2O$

Atmospheric and air-quality monitoring applications require compact, low-cost instruments that provide calibration-free, high-precision data for several molecules and their isotopologues simultaneously. In order to demonstrate the performance that low-power modulator-based dual-comb spectrometers can achieve in such targeted real-world sensing applications, we recorded the normalized cavity-enhanced transmission spectrum of a synthetic air sample. The results are shown in Fig. 4, where the CW carrier wave number was $\tilde{\nu}_0$ = 6 364.678(3) cm$^{-1}$, and the effective interaction path length between the probe comb and the sample was 10 km. Despite the relatively high water content of this synthetic atmospheric sample, the present near-infrared multiplexed approach provided sufficient resolution and sensitivity to yield precise measurements of CO$_2$ and CO mole fractions. The 1σ detection limits for CO$_2$ and CO in the presence of 4.1 % H$_2$O were 13 μmol/mol and 11 μmol/mol, respectively (1 part-per-million (ppm) ≡ 1 μmol/mol). Within this spectral region, both HDO and H$_2$O transitions exist (see Fig. 4). During fitting, the isotopic abundance of HDO was fixed to be 3.107 × 10$^{-4}$ of the entire water content [37].

The noise-equivalent absorption coefficient normalized to one resolution element was NEA = $\alpha\sqrt{T/M}$ = 1.5 × 10$^{-8}$ cm$^{-1}$ Hz$^{-1/2}$, where $\alpha$ is the intensity absorption coefficient and $T$ is the total integration time. The NEA coefficient was calculated from the spectrum shown in Fig. 4, where $\alpha$ = 1.2 × 10$^{-8}$ cm$^{-1}$, $M$ = 200, and $T$ = 320 s. This value is slightly higher than the dual-comb spectroscopy NEA coefficient recently reported by Millot *et al.* [22], but achieved here with more than five orders of magnitude less optical power per comb tooth. Baseline fluctuations, whether due to drifts in electronics, instrument alignment, or interfering broadband absorption, were readily fit along with the narrow molecular absorption features. Thus calibration-free, targeted sensing of multiple species and their isotopologues was realized using extremely low optical powers. Even at a modest optical bandwidth of ~ 50 GHz, the acquisition of a few well-known spectral transitions was sufficient to quantitatively identify the constituents of the given gas mixture. Other applications that study targeted molecular transitions may benefit from the improved acquisition speed and multiplexed nature of robust dual-EOM comb spectroscopy [38, 39].

The synthetic air sample comprises 0.6 kPa of water (H$_2$O and HDO in natural abundances), 0.6 kPa of 2 500 μmol/mol CO in N$_2$, and a balance of room air (total pressure of 13 kPa, or ≈ 100 Torr). Known spectroscopic parameters [40], a superposition of Voigt line profiles, and a polynomial baseline were used to model the cavity-enhanced transmission spectrum of the synthetic air sample. We begin our model with the equation for the cavity-transmitted complex electric field $\tilde{E}_t$ as a function of frequency $\nu$ shown in Eq. (1) [41].

$$\tilde{E}_t(\nu) = \tilde{E}_0 \frac{t_m t'_m \exp\left[-(\alpha(\nu)L + i\Delta\varphi(\nu))/2\right]}{1 - r_m r'_m \exp(-\alpha(\nu)L - i\Delta\varphi(\nu))} \quad (1)$$

In Eq. (1), $\tilde{E}_0$ is the incident electric field, $L$ is the cavity length ($L$ = 73.8 cm), $t_m$ and $t'_m$ are the mirror amplitude transmission coefficients, and $r_m$ and $r'_m$ are the mirror amplitude reflection coefficients. It is a good assumption that $t_m = t'_m$ (and that $r_m = r'_m$), as both supermirrors were from the same coating run. The intensity transmission and reflection coefficients can then be defined as $T_m = (t_m)^2$ and $R_m = (r_m)^2$. Additionally, each mirror is assumed to have identical intrinsic intensity losses $\ell_m$. By conservation of energy, $R_m + T_m + \ell_m = 1$. The intensity absorption coefficient $\alpha$ is related to the real part of the complex lineshape function as shown in Eq. (2).

$$\alpha(\nu) = S n_x \operatorname{Re}(\tilde{g}(\nu)) \quad (2)$$

In Eq. (2), $S$ is the line intensity, $n_x$ is the concentration of molecular species $x$, and $\tilde{g}$ is the complex Voigt function. The dispersion term $\Delta\varphi$ is also shown in Eq. (1), and in the absence of molecular dispersion, mirror dispersion, and any mismatch between $f_{\text{mod}}$ and the cavity free spectral range is equal to integer multiples of $2\pi$ for all comb mode orders. At the SNR observed in this proof-of-principle experiment, the dispersion term can be considered small, and therefore $|\exp(-i\Delta\varphi/2)| \approx |\exp(-i\Delta\varphi)| \approx 1$.

For cavity-enhanced multiheterodyne spectroscopy, the transmitted electric field $\tilde{E}_t$ (and also $\tilde{E}_t$ in the absence of molecular absorption $\tilde{E}_{t,0}$ to be used for normalization) was interfered with the local oscillator field $\tilde{E}_{LO}$ on a photodetector, as opposed to the standard homodyne operation performed in traditional cavity-enhanced Fourier transform spectroscopy [34]. The LO field contained no information about the optical cavity, and the normalized transmission spectrum (absent a proportionality constant and baseline function) is shown in Eq. (3).

$$\frac{I_t(v)}{I_{t,0}(v)} = \frac{T_m \exp(-\alpha(v)L/2)}{1 - R_m \exp(-\alpha(v)L)} \tag{3}$$

As stated above, perturbations to the nominal cavity mode center frequencies caused by dispersion were presumed small and therefore ignored. The optical bandwidth reported here was also presumed small relative to known frequency-dependent changes in cavity mirror $R_m$, $T_m$, and $L_m$ as well as the cavity free spectral range. For a single cavity round-trip, the molecular sample was optically thin ($\alpha(v)L \ll 1$), and therefore the numerator of Eq. (3) could be simplified as follows: $T_m \exp(-\alpha(v)L/2) \approx T_m(1 - \alpha(v)L/2) \approx T_m$. The averaging of successive normalized transmission spectra was done following conversion to $\alpha$ using Eq. (3) in the optically thin limit.

## 6. Conclusion

Given the quasi continuous-wave nature of the optical interferograms shown in Fig. 1(c), electro-optic modulator combs offer a distinct advantage over their mode-locked laser counterparts as the requirement for exceptionally high-dynamic-range linear photodetectors and digitizers is dramatically reduced. This limitation has in the past required dual-comb spectrometers to utilize optical bandpass filters [5] that reduced their instantaneous bandwidth to values nearly commensurate with that reported here. In addition, the contribution of shot noise from photons outside of the relevant optical bandwidth is largely eliminated with the narrower bandwidth electro-optic combs while fully maintaining the advantages of spectral multiplexing. With < 300 comb teeth, $\Delta f_{mod}$ could be large, which enabled spectral acquisition times as short as 6.7 μs ($2T_i = 2/\Delta f_{mod}$), or 600 times faster than the only demonstration of cavity-enhanced dual comb spectroscopy using mode-locked lasers [4]. Equivalently, the maximum acquisition rate is therefore $\Delta f_{mod}/2 = 150$ kHz. In practice, spectral acquisition is limited to rates shorter than $\Delta f_{mod}/2$ due to dead time during optical switching between probe and reference combs. The study of optical arbitrary waveforms [42, 43], fast chemical kinetics [44], and coherent population transfer [45], as well as new methodologies in coherent medical imaging [18, 46], could all benefit from such sensitive and fast spectral acquisition techniques that utilize emerging optical frequency comb technology. Finally, the demonstration of coherent time-domain averaging for > 2 hours, along with the inclusion of tailored radiofrequency waveforms generated from low-cost, passive devices opens new avenues for the commercialization and ruggedization of multiplexed optical sensors without the demanding stabilization and rigid signal sampling requirements of mode-locked laser dual comb spectroscopy.

## Acknowledgments

This work was funded by the National Institute of Standards and Technology (NIST) Greenhouse Gas Measurement and Climate Research Program. We thank Ian Coddington (NIST) for comments and suggestions regarding this paper.